\documentclass[prl,preprint,onecolumn,superscriptaddress,showpacs]{revtex4}

\usepackage{latexsym}
\usepackage{graphicx}
\usepackage{braket}

\begin{document}

\author{Z. Kim}
\author{B. Suri}
\author{V. Zaretskey}
\author{S. Novikov}
\affiliation{Laboratory for Physical Sciences, College Park,
Maryland, 20740} \affiliation{Department of Physics, University of
Maryland, College Park, Maryland, 20742}
\author{K. D. Osborn}
\author{A. Mizel}
\affiliation{Laboratory for Physical Sciences, College Park,
Maryland, 20740}
\author{F. C. Wellstood}
\affiliation{Department of Physics, University of Maryland,
College Park, Maryland, 20742}
\affiliation{Joint Quantum Institute and Center for Nanophysics and Advanced Materials, College Park, Maryland, 20742}
\author{B. S. Palmer}
\affiliation{Laboratory for Physical Sciences, College Park,
Maryland, 20740}

\title{Decoupling a Cooper-pair box to enhance the lifetime to 0.2
ms}
\newcommand{\etal}{\textit{et al.}}
\newcommand{\kb}{k_{B}}

\date{\today}

\begin{abstract}
We present results on a circuit QED experiment in which a separate transmission line is used to address a quasi-lumped element superconducting microwave resonator which is in turn coupled to an Al/AlO$_{\mbox{x}}$/Al Cooper-pair box (CPB) charge qubit. With our device, we find a strong correlation between the lifetime of the qubit and the inverse of the coupling between the qubit and the transmission line. At the smallest coupling we measured a lifetime of the CPB was $T_{1} = 200\ \mu$s, which represents more than a twenty fold improvement in the lifetime of the CPB compared with previous results.  These results imply that the loss tangent in the AlO$_{\mbox{x}}$ junction barrier must be less than about $4\times 10^{-8}$ at 4.5 GHz, about 4 orders of magnitude less than reported in larger area Al/AlO$_{\mbox{x}}$/Al tunnel junctions.
\end{abstract}

 \maketitle

The use of high quality factor superconducting resonators has many
applications in solid-state and atomic physics including microwave
kinetic inductance detectors (MKID)~\cite{DayNature2003} and in
the quantum information sciences in the form of circuit quantum
electrodynamics (cQED)~\cite{WallraffNature2004,
HertzbergNaturePhysics, AndreNaturePhysics2006}. Understanding and
minimizing the sources of energy loss in these systems has a
general technological importance for all of these topics to
improve the sensitivities of MKIDs and coherence times for qubits.
For superconducting qubits, energy loss has been attributed to
various mechanisms, including discrete charge two-level
fluctuators coupled to the
qubit~\cite{SimmondsPRL2004,ZaeillPRB2008}, dielectric
loss~\cite{MartinisPRL2005}, non-equilibrium
quasiparticles~\cite{PalmerPRB2007} and lossy higher order
electromagnetic modes of the electromagnetic field which are
coupled to the qubit~\cite{HouckPRL2008}.

Here, we report the observation of relaxation times in a
Cooper-pair box (CPB) that are one order of magnitude larger than
previously reported. Our design builds on the circuit quantum
electrodynamics (CQED)
approach~\cite{BlaisPRA2004,WallraffNature2004,WallraffPRL2005}:
we coupled the CPB to a resonator and used perturbations of the
resonator frequency to read-out the state of the CPB over one
octave in frequency. In contrast to previous work, however, we
used a lumped element design for the resonator and addressed it
using a separate transmission line. In our experiment, we find a
key reason for obtaining the long lifetimes was decoupling the CPB
from the transmission line.


%

Our CPB consists of a small (100 nm $\times\ 2\ \mu$m $\times\ 30\
$ nm thick) superconducting Al island connected to superconducting
leads by two ultra-small Josephson junctions [see
Fig.~\ref{fig:Fig1}~(c)]. By applying a dc voltage $V_{g}$ that is
capacitively coupled to the island with capacitance $C_{g}^{*}$,
we can change the system's electrostatic
charging energy, and by varying the
magnetic flux through the superconducting loop we can modulate the
critical current $I_{\circ}$ and therefore the Josephson
energy $E_{J}=\hbar I_{\circ}/2e$. Restricting consideration to the two lowest
levels, the Hamiltonian of the CPB can be written as
\begin{equation}
H_{CPB} = \frac{\hbar \omega_{a}}{2} \sigma_{z} \label{eq:HamCPB}
\end{equation}
where $\hbar \omega_{a}= \sqrt{(4E_{c}(1-n_{g}))^2+E_{J}^2}$,
$E_{c}=e^{2}/2C_{\Sigma}$ is the electrostatic charging energy constant,
 $n_{g}= C_{g}^{*}V_{g}/e$ is the
reduced gate voltage, and $\sigma_{z}$ is a Pauli spin matrix.


We coupled our CPB to a quasi-lumped element resonator
[Fig.~\ref{fig:Fig1}~(a)] and measured the CPB at the charge
degeneracy point while it was tuned over one octave in frequency.
When the CPB is coupled to a resonator and the detuning between
the qubit and the resonator ($\Delta = \omega_{a} -
\omega_{r}$) is large compared to the strength of the coupling
$g$ between them, the Hamiltonian for the combined system
is approximately
\begin{equation}
H \cong \hbar \displaystyle
\left(\omega_{r}+\frac{g^2}{\Delta}\sigma_{z}\right)\left(a^{\dag}a+\frac{1}{2}\right)+\frac{\hbar
\omega_{a}}{2} \sigma_{z}, \label{eq:Htot}
\end{equation}
where $\hbar g=(e C_{g}/C_{\Sigma})\sqrt{\hbar \omega_{r}/2C}$,
$C$ is the capacitance of the resonator with resonance frequency
$\omega_{r}$, and $C_{g}$ is the capacitance between the resonator
and the island of the CPB~\cite{BlaisPRA2004,WallraffPRL2005}.
Depending on the state of the qubit, Eq.~(\ref{eq:Htot}) predicts that
the bare resonance frequency $\omega_{r}$ is shifted by $\pm
g^{2}/\Delta$. For $g/2\pi = 5$ MHz and $\Delta/2\pi = 1$ GHz,
we find the maximum dispersive frequency shift of the resonator's resonance
frequency is $g^2/2\pi \Delta = 25$ kHz.

To measure these small frequency shifts we have designed and
fabricated, using photolithographic lift-off techniques, a high-$Q$
superconducting resonator made from a 100 nm thick film of Al on a
c-plane sapphire wafer. The resonator consists of a coplanar
meander-line inductor ($\sim 2$ nH) and interdigital capacitor
($\sim 400$ fF) coupled to a transmission line [see
Fig.~\ref{fig:Fig1}~(a)].
The resonance frequency of our resonator was $\omega_{r}/2\pi =
5.44$ GHz, the loaded quality factor was $Q_{L} = 22,000$, and the
internal quality factor was $Q_{i}=32,000$. Subsequently, the CPB
was fabricated using e-beam lithography and double-angle
evaporation (with an oxidation in between the two evaporations) to
form the small Josephson junctions~\cite{Fulton1987}. We used a
bilayer of MMA-MMA copolymer and ZEP520 as the electron beam
resist and the 30 nm thick Al island and 50 nm thick Al leads were
deposited in an electron beam evaporator [see
Fig.~\ref{fig:Fig1}~(b)-(c)].

The device was packaged in an rf-tight Cu box and bolted to the
mixing chamber of an Oxford Instruments Kelvinox 100 dilution
refrigerator. To reduce Johnson/Nyquist noise from higher
temperatures, we used cold attenuators on the input microwave line
and two isolators on the output line [see
Fig.~\ref{fig:Fig1}~(d)].  The input microwave
power had 10 dB of attenuation at 4 K, 20 dB at 0.7 K, and 30
dB on the mixing chamber at 25 mK. On the output line, both
isolators on the mixing chamber had a minimum isolation of 18 dB
between 4 and 8 GHz. The output microwave signal was amplified
with a HEMT amplifier sitting in the He bath. To allow a dc gate
voltage bias to be applied to the island of the CPB from the
transmission line, a bias tee was placed on the transmission line
before the device and a dc block was placed on the transmission
line after the device [see Fig.~\ref{fig:Fig1}~(d)].

Figure~\ref{fig:Fig2}~(a) shows a plot of the transition spectrum
of the CPB qubit. This spectrum was taken by measuring the phase
of the transmitted microwaves at the resonator's bare resonance
frequency ($f_{r} = 5.44$ GHz) while sweeping the dc gate voltage,
and stepping the frequency of a second microwave source from 6.2
to 8.4 GHz. When the second microwave source is resonant with the
transition between the two lowest states of the CPB, the CPB is
excited. This causes a change in $\omega_{r}$ [see
Eq.~(\ref{eq:Htot})] and a change in the phase of the transmitted
signal. For these measurements the average number of photons in
the resonator was $\overline{n} = 20$ photons. From fitting this
spectrum, we extract $E_{c}/h = 6.24$ GHz and $E_{J}/h = 6.35$
GHz. Using these parameters and the measured dispersive shift
($g^2/2\pi \Delta \simeq 27$ kHz), we extracted the coupling between
the resonator and the CPB, $g/2\pi = 5$ MHz.

To measure Rabi oscillations, we applied magnetic flux to set $E_{J}/h=6.15$ GHz, dc biased
the gate voltage at the charge degeneracy point
$n_{g} = 1$ and delivered a short pulse of microwaves at $f =
6.15$ GHz while continuously monitoring the phase of the resonator
with an average of $\overline{n} = 20$ photons.
Figure~\ref{fig:Fig2}~(b) shows a false color plot of the measured
phase (which has been calibrated in terms of the probability of
occupancy of the excited state) as a function of time after
sending the pulse and
as a function of the length of the pulse.
Fig.~\ref{fig:Fig2}~(c) presents a linecut
through~\ref{fig:Fig2}~(b); we see clear driven oscillations of
the state of the qubit.


Figure~\ref{fig:Fig2}~(d) shows a plot of the probability $P_{e}$
of occupying the excited state as a function of time after sending
a $\pi$ pulse to the qubit at $f = 6.15$ GHz and $n_{g} = 1$. For
$P_{e}>5~\%$, the relaxation is well fit by an exponential with a
decay time of $T_{1} = 30\ \mu$s.  We also varied the Josephson
energy from a maximum of $E_{J}/h = 19$ GHz and measured $T_{1}$
at the charge degeneracy point over one octave in the CPB
transition frequency, from 3.8 to 8.5 GHz [black squares in
Fig.~\ref{fig:Fig3}~(b)].  While $T_{1} \sim 30\ \mu$s for
frequencies above $f_{r}$, we discover that the CPB attains a
striking lifetime of $T_{1} = 200\ \mu$s below $f_{r}$ at $f =
4.5$ GHz.

Some of the qualitative features in Fig.~\ref{fig:Fig3}~(b) can be
understood. In particular, the depressions in $T_{1}$ at $f =
4.18$ GHz and $f=5.67$ GHz correlate to changes in the measured
transmission of microwaves through the transmission line [see
Fig.~\ref{fig:Fig3}~(a)] and are likely due to the packaging of
our device ($f=4.18$ GHz) or imperfections in a microwave
component ($f=5.67$ GHz). Also, the dip near $f_{r} = 5.44$ GHz is
consistent with enhanced spontaneous emission at the resonator
frequency due to the Purcell effect [see dashed blue curve in
Fig.~\ref{fig:Fig3}~(b)]~\cite{HouckPRL2008}.

Next, we studied the coupling between the qubit and the microwave
drive to understand the steady change in $T_{1}$ below $f_{r}$. At
several values of $f$, we measured the change in the frequency
$f_{\mathrm{Rabi}}$ of the Rabi oscillations with microwave drive
voltage $V$. The red triangles in Fig.~\ref{fig:Fig3}~(b) show
$dV/df_{\mathrm{Rabi}}$ versus $f$. This quantity indicates how
decoupled the transmission line is from the qubit; when
$dV/df_{\mathrm{Rabi}}$ is  large, the qubit responds only weakly
to a change in $V$ and when $dV/df_{\mathrm{Rabi}}$ is small, the
qubit responds strongly to a change in $V$.  While the simple
model for our system [Fig.~\ref{fig:Fig1}~(d)] does not predict
this behavior of the coupling we note that the coupling is
changing near and between additional resonances in the system
which can produce a non-trivial dependence of
$dV/df_{\mathrm{Rabi}}$ on $f$. We find that we can achieve good
agreement between the experimental $dV/df_{\mathrm{Rabi}}$ and a
theoretical calculation that augments the simple circuit of
Fig.~\ref{fig:Fig1}~(d) with an additional LC circuit which is
coupled to the transmission line and the qubit to model the
microwave packaging resonance at $f=4.18$ GHz.  



A close relationship between $T_1$ and the decoupling $dV/df_{\mathrm{Rabi}}$ is evident in the figure.  
If we assume that the qubit is capacitively coupled to a $Z_{\circ} = 50\ \Omega$ quantum dissipative environment at the input and output microwave lines,  then the
decay rate is given by~\cite{SchoelkopfQuantumNoise2003}
\begin{equation}
T_{1}^{-1}= \left(\frac{df_{\mathrm{Rabi}}}{dV}\right)^{2} 8
\pi^{2} Z_{\circ} h f. \label{eq:T1chargeVnoise}
\end{equation}
The filled diamonds in Fig.~\ref{fig:Fig3}~(b) show that
Eq.~(\ref{eq:T1chargeVnoise})  with an additional unknown fixed
decay rate of $T_{1}^{-1} = 5\times 10^{3}$ s$^{-1}$ is in
reasonably good qualitative agreement with the data (filled
squares).  This relationship suggests that decoupling the qubit from the noisy
transmission line in our experiment was
essential to allowing $T_1$ to reach  $30\ \mu$s at most values of
$f$ and to attain $200\ \mu$s at $f = 4.5$ GHz.






%

The measured lifetime also places a bound on charge noise in the
CPB. If charge noise is the dominant mechanism producing
relaxation then the spectral density of charge noise $S_{Q}$ at
positive frequencies is related to $T_{1}$ at the charge
degeneracy point
by~\cite{SpecDenNoise,AstafievPRL2004,SchoelkopfQuantumNoise2003},
\begin{equation}
S_{Q}(+f) = \left(\frac{e\hbar}{2E_{c}}\right)^{2}\frac{1}{T_{1}}.
\label{eq:chargenoise}
\end{equation}
Using Eq.~(\ref{eq:chargenoise}) and the measured value of $T_{1}$ at 4.5 GHz, we get an upper bound on the spectral
density of charge noise of $S_{Q} (f = 4.5\ \mathrm{GHz}) \leq
10^{-18}$ $e^{2}$/Hz. This level of charge noise is approximately
an order of magnitude smaller than the bound measured by Vion
\etal~\cite{VionScience2002}. If we assume that $S_{Q}$ has a
$1/f$ dependence, then the symmetrized \emph{classical} spectral
density of charge noise at 1 Hz would be approximately $S_{Q}(f =
1\ \mathrm{Hz}) = 2 (10^{-4})^{2}$ $e^{2}$/Hz, a value that is two
orders of magnitude smaller than is typically measured at low
frequencies~\cite{ZimmerliAPL1992,KenyonJAP2000} and similar to
the best values reported in stacked SETs~\cite{KrupeninJAP1998}.

Our $T_{1}$ measurements also place a bound on dielectric loss in the Josephson junctions.
If $T_{1}$ is limited by dissipation in the junction, then the effective resistance of
the tunnel junctions is related to the charge noise $S_{Q}$ by
\begin{equation}
R = \frac{2\hbar}{\omega S_{Q}}
=\frac{T_{1}}{C_{\Sigma}}\left(\frac{4E_{c}}{E_{J}}\right).
\end{equation}
At $E_{J}/h$ = 4.5 GHz, where $T_{1} = 200\ \mu$s, this yields $R
\sim 3 \times 10^{11}$ Ohms. If this dissipation were due to
dielectric loss in the amorphous AlO$_{\mbox{x}}$ tunnel junction
barrier, then one would find $\tan \delta =(R \omega
C_{\Sigma})^{-1} = 4 \times 10^{-8}$ which appears to be four
orders of magnitude smaller than most amorphous dielectrics at
both low temperatures and low microwave
powers~\cite{MartinisPRL2005}. A possible explanation is that the
loss is due to a few discrete two-level fluctuators (TLFs) in the
ultra-small junctions. Spectroscopic measurements on CPB devices
have shown anomalous avoided level crossings with splitting sizes
on the order of 50 MHz and decay rates due to the TLF on the order
of 10 $\mu$s$^{-1}$~\cite{ZaeillPRB2008}. If we take these
parameters and assume that the TLF resonance is detuned by 2 GHz
from the CPB resonance, then the $T_{1}$ from a single fluctuator
would be approximately 160 $\mu$s.


Another metric of charge noise can be found from dephasing
measurements. To minimize dephasing from photons in the
resonator~\cite{BlaisPRA2004}, the power at $f_{r}$ was pulsed on
only after the state of the CPB was manipulated. At
$E_{J}/h = 6.4$ GHz we find a Ramsey decay time of $T_{2}^{*} =
70$ ns. Assuming 1/$f$ charge noise is the dominant free induction
dephasing mechanism~\cite{AstafievPRL2004}, then at $n_{g}=1$ the
standard deviation of the charge noise ($\sigma_{Q}$)
obeys~\cite{AstafievPRL2004}
\begin{equation}
\sigma_{Q}^{2} =
\frac{1}{T_{2}^{*}}\frac{E_{J}}{(4E_{c})^{2}}\frac{2e^{2}\hbar}{\eta}
\label{eq:SqRabi}
\end{equation}
where $\eta = ln(f_{max}/f_{min})$ and $f_{min}$ and $f_{max}$ are
the minimum and maximum  bandwidth of the measurement, respectively. Using Eq.~(\ref{eq:SqRabi}) we find $\sigma_{Q} =
(2\times 10^{-3}\ e)^{2}$ which is a fairly typical value for the
amplitude of $1/f$ charge
noise~\cite{ZimmerliAPL1992,KenyonJAP2000}. Measurements of the
decay of Rabi oscillations showed a maximum decay time of $T'
\simeq 1\ \mu$s.

We also obtained some measurements on a second device with a charging
energy of $E_{c}/h = 12.48$ GHz. The lifetime of that device at $f =
E_{J}/h = 7.5$ GHz was $T_{1} = 8\ \mu$s which from Eq.~(\ref{eq:chargenoise}) gives $S_{Q}(f = 7.5\ \mathrm{GHz})
= 5 \times 10^{-18}\ e^{2}$/Hz. This value is within a factor of
two of the device discussed in this paper at $f = 7.5$ GHz.
Unfortunately, we did not obtain $T_{1}$ measurements on this second device over a wide range of frequency before the device stopped functioning.

In conclusion, we have measured the spectrum, excited state
lifetime, and Rabi oscillations of a CPB qubit over one octave in
transition frequency. We find $T_{1}$ varies from 4 $\mu$s at $f=
8$ GHz up to 200 $\mu$s at $f= 4.5$ GHz.     The longest lifetime
places an upper bound on the spectral density of charge noise
which is $S_{Q} (f = 4.5\ \mathrm{GHz}) \leq 10^{-18}$ $e^{2}$/Hz
at 4.5 GHz.  Our measurements place a remarkably small upper bound
on dielectric loss in the junction barrier.
While the exact source of improvement in the lifetime of our CPB compared with other results~\cite{WallraffPRL2005,VionScience2002,AstafievPRL2004} is unknown, our measurements suggest that the coupling between the qubit and the transmission line can play a key role.




\acknowledgments{FCW would like to acknowledge support from the
Joint Quantum Institute and the Center for Nanophysics and
Advanced Materials. The authors would like to acknowledge
discussions with Daniel Braun, David Schuster, and Andrew Houck.}

\clearpage
\begin{figure}
\caption{(a) Optical image of quasi-lumped element resonator coupled to transmission line and surrounded by ground plane. White regimes are aluminum and black regimes are sapphire. (b) Optical image of CPB close to the
interdigital capacitor. (c) Scanning electron micrograph of CPB. (d) Schematic of the measurement set-up.
Two microwave tones are sent to the device on the mixing chamber
through microwave lines and attenuators at different temperatures.
On the mixing chamber the microwave tones are combined
with a dc voltage before the device. After the device
the signal passes through two isolators, is amplified at both
4 and 300 K, mixed to a smaller intermediate frequency and then digitized on an oscilloscope.
}
\label{fig:Fig1}
\end{figure}

\begin{figure}
\caption{(a) Measured spectrum of CPB. The gray scale plot shows
the change in phase of the transmitted microwaves at the probe frequency as a function of the
pump frequency and $n_{g}$. (b) Rabi oscillation of CPB qubit for
microwave drive at $f = 6.15$ GHz. (c) Line cut of (b) along the
pulse length at a measurement time of 2 $\mu$s. The maximum
measured population in the excited state was about 80 \%. From the
fit (red curve), the extracted Rabi frequency was 39 MHz. (d)
Energy relaxation measurement of CPB from the excited state.
Red line shows fit with $T_{1}\simeq 30~\mu$s.} \label{fig:Fig2}
\end{figure}

\begin{figure}
\caption{(a) Plot of the ratio of the transmitted output voltage
before the mixer in Fig.~\ref{fig:Fig1} to input voltage
($S_{21}$) versus frequency through the system. The arrow at 5.44
GHz identifies the resonance of the resonator. (b) Log plot of
measured $T_{1}$ versus frequency (filled squares) and model for
$T_{1}$ (filled diamonds) based on the measured coupling to
quantum noise from 50 $\Omega$. Dashed blue curve shows
contribution to loss from coupling to resonator plus an additional
decay rate of $T_{1}^{-1} = 5 \times 10^{3}$ s$^{-1}$ below
$f_{r}$ and $T_{1}^{-1} = 2 \times 10^{4}$ s$^{-1}$ above $f_{r}$.
Right axis: Inverse of the measured coupling (Rabi frequency
divided by applied rms voltage) between the transmission line and
the CPB versus $f$ (red filled triangles). } \label{fig:Fig3}
\end{figure}

\clearpage
\begin{figure}[t]
\centering \includegraphics[width=0.8\textwidth,angle=0]{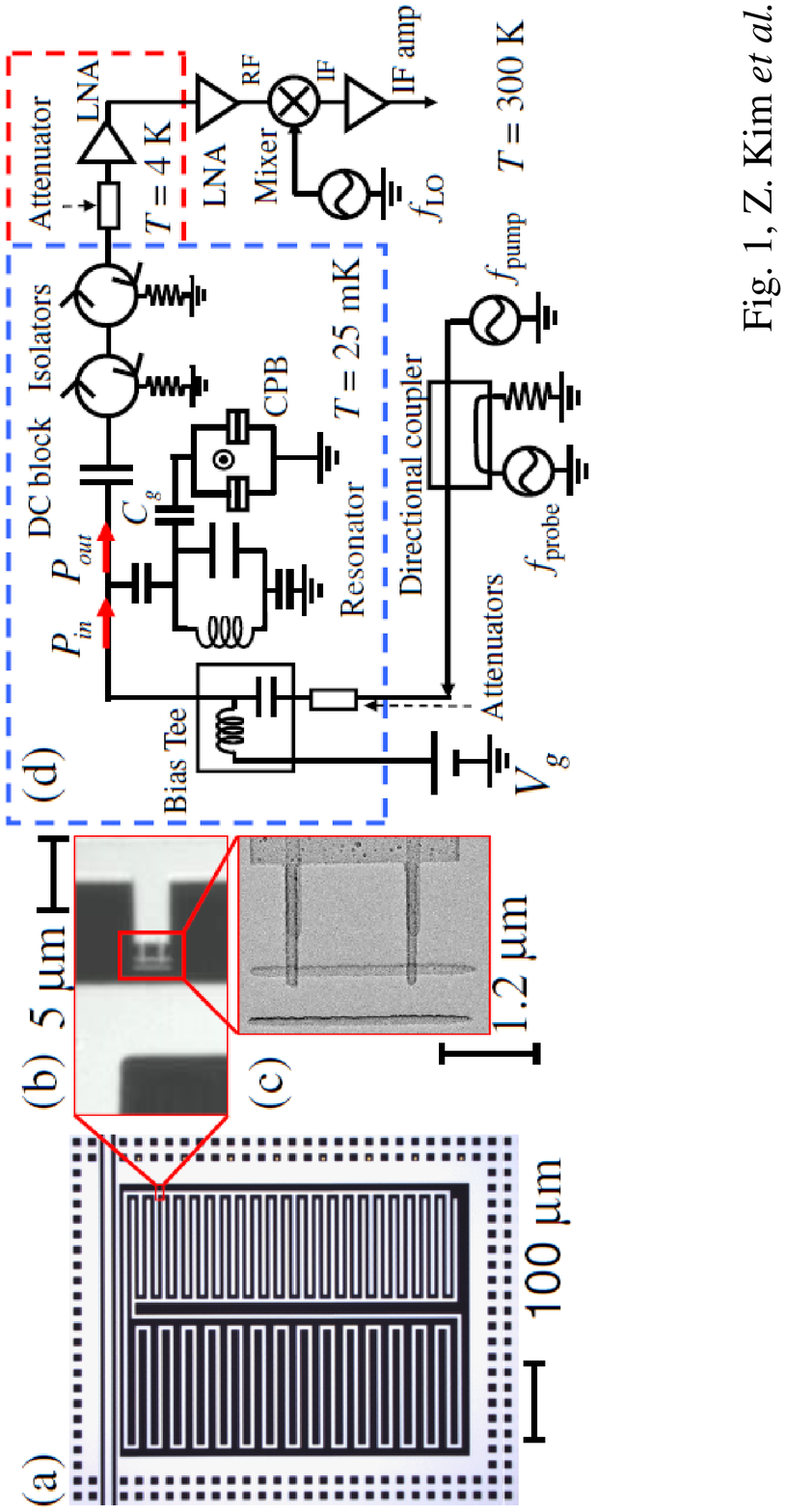}
\end{figure}

\clearpage
\begin{figure}[t]
\centering \includegraphics[width=1.1\textwidth,angle=0]{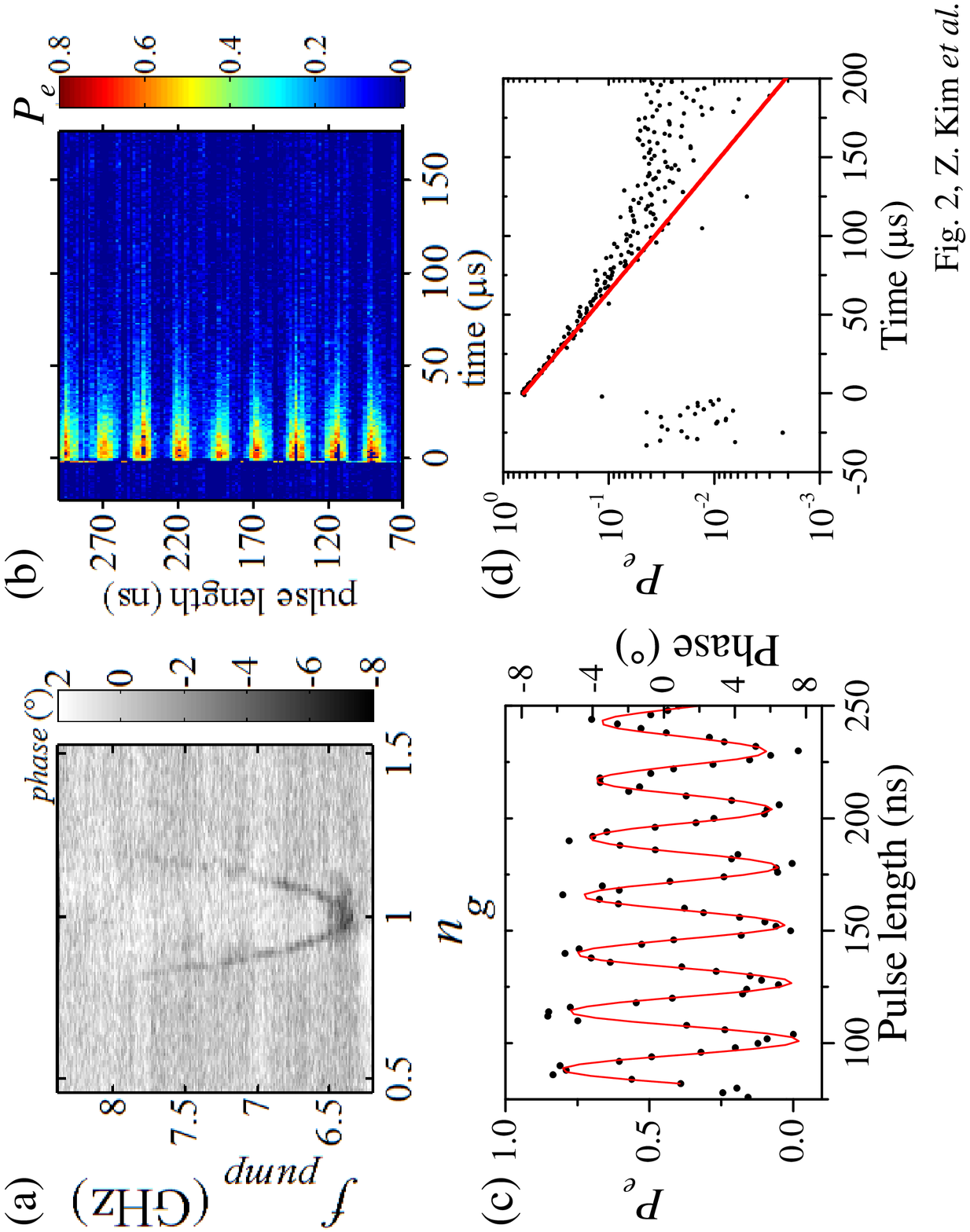}
\end{figure}

\clearpage
\begin{figure}[t]
\centering \includegraphics[width=1.1\textwidth,angle=0]{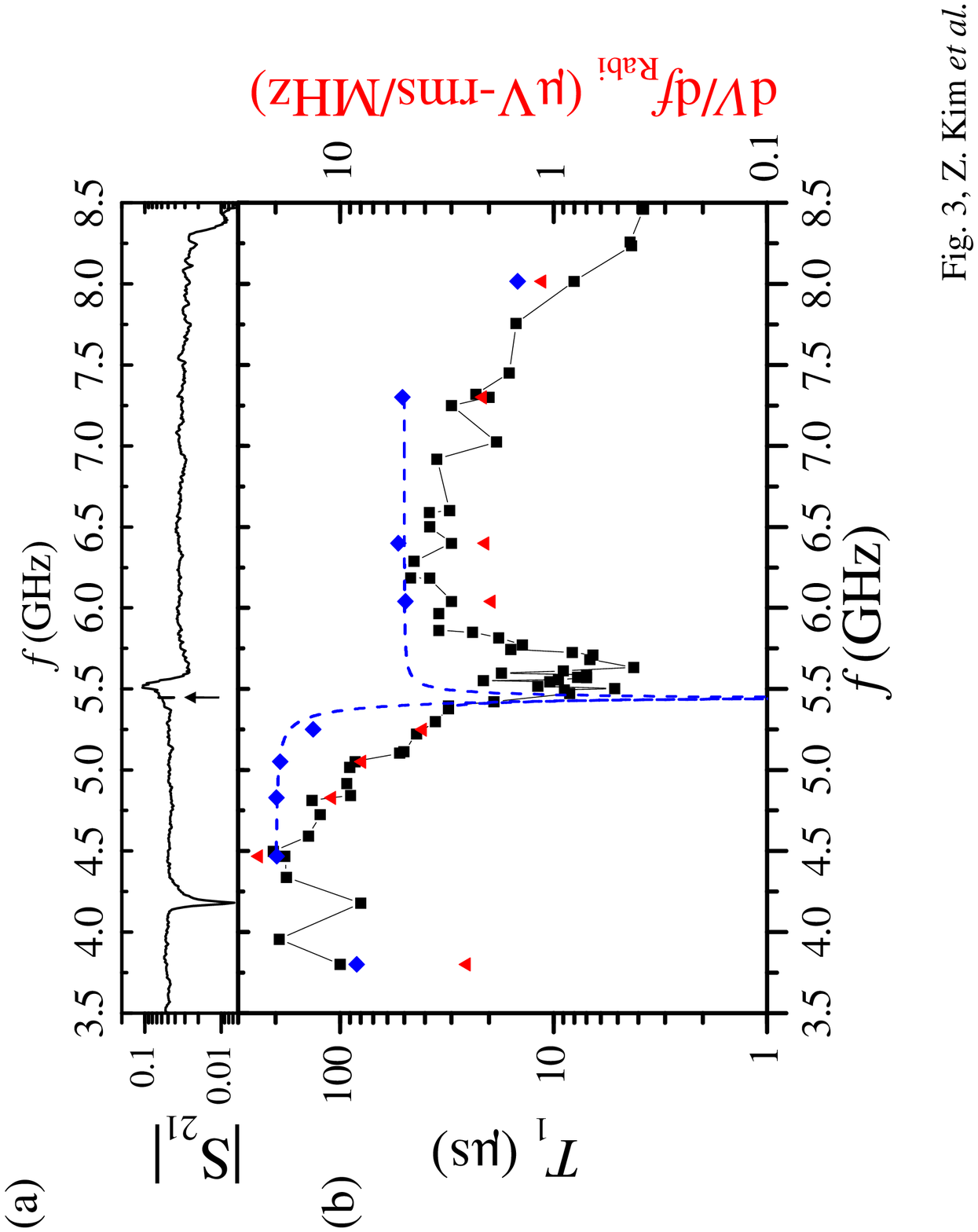}
\end{figure}
\clearpage

\begin{thebibliography}{10}
%
%
%
%
%
\bibitem{DayNature2003} P. K. Day, H. G. LeDuc, B. A. Mazin, A.
Vayonakis, and J. Zmuidzinas, Nature \textbf{425}, 817 (2003).
%
\bibitem{WallraffNature2004} A. Wallraff  \etal, Nature \textbf{431}, 162 (2004).
%
\bibitem{HertzbergNaturePhysics} J. Hertzberg \etal, Nature
Physics \textbf{6}, 213 (2010).
%
\bibitem{AndreNaturePhysics2006} A. Andr\'{e} \etal, Nature Physics
\textbf{2}, 636 (2006).
%
%
\bibitem{SimmondsPRL2004} R. W. Simmonds \etal, Phys. Rev. Lett. \textbf{93},
077003 (2004).
%
\bibitem{ZaeillPRB2008} Z. Kim \etal, Phys. Rev. B \textbf{78}, 144506 (2008).
%
\bibitem{MartinisPRL2005} J. M. Martinis \etal, Phys. Rev. Lett.
\textbf{95}, 210503 (2005).
%
\bibitem{PalmerPRB2007} B. S. Palmer \etal, Phys. Rev. B \textbf{76}, 054501 (2007).
%
\bibitem{HouckPRL2008} A. A. Houck \etal, Phys. Rev. Lett. \textbf{101}, 080502 (2008).
%
\bibitem{BlaisPRA2004} Alexandre Blais, Ren-Shou Huang, Andreas Wallraff, S. M. Girvin, and R. J. Schoelkopf, Phys. Rev. A \textbf{69}, 062320 (2004).
%
\bibitem{WallraffPRL2005} A. Wallraff \etal, Phys. Rev. Lett. \textbf{95}, 060501 (2005).
%
\bibitem{AstafievPRL2004} O. Astafiev , Y. A. Pashkin, Y. Nakamura, T. Yamamoto, and J. S. Tsai Phys. Rev. Lett. \textbf{93}, 267007 (2004).
%
\bibitem{VionScience2002} D. Vion \etal, Science \textbf{296}, 886 (2002).
%
\bibitem{Fulton1987} T. A. Fulton and G. J. Dolan, Phys. Rev. Lett. \textbf{59}, 109 (1987).
%
\bibitem{SchoelkopfQuantumNoise2003} R. J. Schoelkopf \etal, in Quantum
Noise in Mesoscopic Physics, edited by Y. V. Nazarov (Kluwer
Academic Publishers, Netherlands, 2003) pp. 175.
%
\bibitem{SpecDenNoise} The spectral density of noise of operator $\hat{Q}$ is defined as
$S_{Q}(f) = \int_{-\infty}^{\infty}e^{i2\pi f
\tau}\braket{\hat{Q}(\tau)\hat{Q}(0)}d\tau$.
%
\bibitem{ZimmerliAPL1992} G. Zimmerli, T. M. Eiles, R. L. Kautz, and J. M. Martinis, Appl. Phys.
Lett. \textbf{61}, 237 (1992).
%
\bibitem{KenyonJAP2000} M. Kenyon, C. J. Lobb, and F. C. Wellstood, Jour. of Appl. Phys. \textbf{88}, 6536
(2000).
%
\bibitem{KrupeninJAP1998} V. A. Krupenin \etal, J. Appl. Phys. \textbf{84}, 3212 (1998).
%
\end{thebibliography}
\end{document}